\documentclass[12pt]{article}
\usepackage{amsmath,amssymb,amsthm,amsxtra,bbm,overpic,bm,epsfig,subfigure,tikz,tikz-feynman}
\usepackage{hyperref}
\usepackage{mathrsfs}
\usepackage{enumitem}
\usepackage{graphicx}
\usepackage{color}
\usepackage{comment}
\usepackage{epstopdf}
\usepackage{float}
\usepackage{cite}
\allowdisplaybreaks[2]
\usetikzlibrary{arrows.meta}
\makeatletter
\tikzfeynmanset{compat=1.1.0} % 启用兼容模式（适用于 pdflatex）
\makeatother
\usepackage[all]{xy}
\usepackage{slashed,stmaryrd,multirow}
 \numberwithin{equation}{section}

\def\thefootnote{\fnsymbol{footnote}}

\usepackage{hyperref}
\usepackage{slashed,stmaryrd,orcidlink}

\usepackage{lscape}%
\usepackage{array}
\usepackage{booktabs}%

\begin{document}
	
	\vspace{0.2cm}
	
	\begin{center}
		{\Large\bf Probing unitarity violation of lepton flavor mixing matrix with reactor antineutrinos at JUNO and TAO}
	\end{center}
	
	\vspace{0.2cm}
	
	\begin{center}
		{\bf Jihong Huang}~{\orcidlink{0000-0002-5092-7002}},$^{1,2}$~\footnote{E-mail: huangjh@ihep.ac.cn}
		\quad
		{\bf Shun Zhou}~{\orcidlink{0000-0003-4572-9666}}~$^{1,2}$~\footnote{E-mail: zhoush@ihep.ac.cn}
		\\
		\vspace{0.2cm}
		{\small
			$^{1}$Institute of High Energy Physics, Chinese Academy of Sciences, Beijing 100049, China\\
			$^{2}$School of Physical Sciences, University of Chinese Academy of Sciences, Beijing 100049, China}
	\end{center}

	\vspace{0.5cm}
	
	\begin{abstract}
Motivated by the precise measurements of neutrino oscillation parameters at Jiangmen Underground Neutrino Observatory (JUNO), we investigate the possibility of probing the unitarity violation of lepton flavor mixing matrix solely with reactor antineutrinos. First, we stress that it is necessary to reconsider the production and detection of neutrinos in a self-consistent way, apart from neutrino propagation, when analyzing experimental sensitivities to unitarity violation. Then, concentrating on JUNO and its satellite experiment Taishan Antineutrino Observatory (TAO), we demonstrate how the event rates of inverse beta decays (i.e., $\overline{\nu}^{}_e + p \to e^+ + n$) for observing $\overline{\nu}^{}_e \to \overline{\nu}^{}_e$ oscillations, and those of elastic antineutrino-electron scattering (i.e., $\overline{\nu}^{}_\alpha + e^- \to \overline{\nu}^{}_\alpha + e^-$ with $\alpha = e, \mu, \tau$) for $\overline{\nu}^{}_e \to \overline{\nu}^{}_\mu$ and $\overline{\nu}^{}_e \to \overline{\nu}^{}_\tau$ oscillations, depend on the parameters characterizing unitarity violation. Our investigation will be useful for JUNO and TAO to place independent constraints with more data in the near future.
	\end{abstract}

	\def\thefootnote{\arabic{footnote}}
	\setcounter{footnote}{0}
	
%	\newpage

	\newpage

\section{Introduction}

Recently, the Jiangmen Underground Neutrino Observatory (JUNO)~\cite{JUNO:2015zny} releases its first results of neutrino oscillation parameters, based on the data accumulated over 59.1 days since its completion in August this year. Even with such a short period of operation, JUNO has already provided us with the most precise measurements of two relevant oscillation parameters~\cite{JUNO:2025gmd}:
\begin{eqnarray}
    \sin^2\theta_{12}^{} = 0.3092 \pm 0.0087 \;, \qquad \Delta m_{21}^2 = \left(7.50 \pm 0.12\right) \times 10^{-5}~{\rm eV}^2 \;,
\end{eqnarray}
reducing the relative uncertainty to $2.81\%$ and $1.55\%$, respectively. Together with its satellite experiment Taishan Antineutrino Observatory (TAO)~\cite{JUNO:2020ijm}, which will start data-taking very soon, JUNO is able to determine the neutrino mass ordering and perform high-precision measurements of the oscillation parameters $\theta_{12}^{}$, $\Delta m_{21}^2$ and $\Delta m_{31}^2$ at the sub-percent level~\cite{JUNO:2022mxj,Capozzi:2025wyn}. 

Since neutrinos are massless in the Standard Model (SM) of elementary particles, new physics beyond the SM is required to accommodate the observed neutrino oscillation phenomena~\cite{ParticleDataGroup:2024cfk,Xing:2020ijf}. In the minimal extension of the SM with an effective Majorana mass matrix for three active neutrinos, the lepton flavor mixing matrix turns out to be unitary and is conventionally parametrized in terms of three mixing angles $\{\theta^{}_{12}, \theta^{}_{13}, \theta^{}_{23}\}$, one Dirac-type CP-violating phases $\delta_{\rm CP}^{}$ and two Majorana-type CP-violating phases $\{\rho, \sigma\}$. Neutrino oscillation experiments are only sensitive to three mixing angles, the Dirac CP-violating phase and two mass-squared differences $\Delta m^2_{ji} \equiv m^2_j - m^2_i$ (for $ji = 21, 31$), where $m^{}_i$ (for $i = 1, 2, 3$) denote three neutrino masses. With the precision measurements of neutrino oscillation parameters, there are two different ways to test the unitarity of lepton flavor mixing matrix, depending on the theoretical framework of neutrino mass generation and lepton flavor mixing. 
\begin{itemize}
  \item The first way is to assume in the first place the unitarity of lepton flavor mixing matrix and then measure neutrino oscillation parameters in different experiments as precisely as possible. In this case, one can verify the unitarity conditions based on independent measurements of all the mixing angles and the Dirac-type CP-violating phase.
  \item The second way is to adopt the complete theory of neutrino masses and lepton flavor mixing, where the unitarity of lepton flavor mixing matrix is indeed violated, and carefully analyze the production, propagation and detection of neutrinos. This is important as the underlying new physics can lead to nonnegligible modifications even to the SM processes.
\end{itemize}
 
In this paper, we shall follow the second approach and reconsider the question of whether the reactor antineutrino experiments, such as JUNO and TAO, are sensitive to the unitarity violation. On the theoretical side, there are well-motivated models for naturally generating tiny neutrino masses, e.g., the type-I~\cite{Minkowski:1977sc,Yanagida:1979as,Gell-Mann:1979vob,Glashow:1979nm,Mohapatra:1979ia} and type-III seesaw models~\cite{Foot:1988aq}, where the $3\times 3$ lepton flavor mixing matrix associated with the charged-current (CC) interactions of the SM leptons is intrinsically non-unitary. Although this question has been extensively studied in previous works, the present paper differs from those in at least two aspects. First, apart from the CC interactions of reactor antineutrinos in the production and detection processes, we also examine the elastic antineutrino-electron scattering (i.e., $\overline{\nu}^{}_\alpha + e^- \to \overline{\nu}^{}_\alpha + e^-$ with $\alpha = e, \mu, \tau$) for observing reactor antineutrinos, where both CC and neutral-current (NC) interactions are involved. For the detection of neutrinos via the NC interactions, it remains confusing from the existing literature how neutrino oscillation probabilities should be defined. Second, we examine quantitatively how the event rates of reactor antineutrinos are calculated and measured, so as to clarify whether or not reactor antineutrino experiments are sensitive to unitarity violation. It is interesting to notice that the answer relies on how the input parameters are chosen and how the data analysis is done.

The remaining part of this paper is organized as follows. In Sec.~\ref{sec:prob}, we derive the formulas of neutrino oscillation probabilities with a non-unitary mixing matrix for both disappearance and appearance channels. Explicit expressions are given in the lower-triangular parametrization of the mixing matrix. The event rates for the inverse beta decays (IBD) at TAO and elastic scattering with electrons at JUNO are calculated in Sec.~\ref{sec:event}, where the sensitivities to the non-unitarity parameters are studied in detail. Some comments on the comparison between our results and those in the literature are also made. Finally, we summarize our main conclusions in Sec.~\ref{sec:sum}.

\section{Neutrino oscillations without unitarity}

\label{sec:prob}

Neutrino flavor eigenstates $\left| \nu^{\rm P}_\alpha \right>$ and $\left| \nu^{\rm D}_\alpha \right>$ (for $\alpha = e, \mu, \tau$) can be defined according to their production and detection processes via the CC interactions, where a specific flavor of charged leptons $l_\alpha^\pm$ is also present. In general, the definition of neutrino flavor eigenstates depends on the production and detection processes~\cite{Giunti:2008cf}. However, since neutrino masses are negligibly small compared to the characteristic energy scales of neutrino production and detection, the definition becomes universal whenever neutrino masses can be neglected, i.e.,
\begin{eqnarray}
  \left| \nu^{}_\alpha \right> \equiv \sum^3_{i=1} N^*_{\alpha i} \left| \nu^{}_i \right> 
 \;,
\end{eqnarray}
where the lepton flavor mixing matrix $N$ is assumed to be unitary (i.e., $NN^\dagger = N^\dagger N = {\bf 1}$) and $\left| \nu^{}_i \right>$ (for $i = 1, 2, 3$) refer to three neutrino mass eigenstates. It has been demonstrated in Ref.~\cite{Giunti:2008cf} that the production and detection rates of neutrinos can be calculated in the same manner as in the SM with massless neutrinos, while the neutrino propagation can be handled separately. The probability amplitudes of neutrino oscillations $\nu^{}_\alpha \to \nu^{}_\beta$ are computed in the usual way by projecting the spacetime-evolved neutrino flavor eigenstate $\left| \nu^{}_{\alpha}(t, L)\right>$ onto the flavor state $\left| \nu^{}_\beta \right>$ at the detector, where $t$ and $L$ stands for the propagation time and distance (with $t\approx L$ for relativistic neutrinos), respectively.

When the $3 \times 3$ lepton flavor mixing matrix $N$ is non-unitary, the neutrino flavor eigenstates $\left|\nu_\alpha^{}\right>$ (for $\alpha=e,\mu,\tau$) can be similarly defined as
\begin{eqnarray}
	\label{eq:flavor_state}
	\left|\nu_\alpha^{}\right> \equiv \sum^3_{i=1} \frac{N_{\alpha i}^{*}}{\sqrt{\left(N N^\dagger\right)_{\alpha\alpha}^{}}} \left|\nu_i^{}\right> \;,
\end{eqnarray}
where the neutrino mass eigenstates are orthonormal, i.e., $\left<\nu_j^{} | \nu_i^{} \right>=\delta_{ij}^{}$, but the flavor eigenstates are not orthogonal any more. However, if neutrino masses can be neglected in the production and detection processes, the latter are treated as in the SM but now with a non-unitary lepton flavor mixing matrix. The factor $\sqrt{\left(N N^\dagger\right)_{\alpha\alpha}^{}}$ in the denominator is introduced to properly normalize the oscillation probability. One can solve the Schr\"{o}dinger-like equations for the time-evolved neutrino flavor eigenstates $\left|\nu_\alpha^{}(t)\right>$. As a result, the probability of observing the flavor eigenstate $\left|\nu_\beta^{}\right>$ at the distance $t \approx L$ for an incident neutrino beam in the flavor state $\left|\nu_\alpha^{}\right>$ from the source is~\cite{Antusch:2006vwa,Xing:2007zj}
\begin{eqnarray}
	\label{eq:prob}
	P_{\alpha\beta}^{} &=& \frac{1}{\left(N N^\dagger\right)_{\alpha\alpha}^{} \left(N N^\dagger\right)_{\beta\beta}^{}} \left[\left|\left(N N^\dagger\right)_{\alpha\beta}^{}\right|^2 - 4 \sum_{i<j} {\rm Re} {\cal N}_{\alpha\beta}^{ij} \sin^2 F^{}_{ji} + 2 \sum_{i<j} {\rm Im} {\cal N}_{\alpha\beta}^{ij} \sin2F^{}_{ji} \right] \;, \quad
\end{eqnarray}
with ${\cal N}_{\alpha\beta}^{ij} \equiv N_{\alpha i}^{} N_{\beta j}^{} N_{\alpha j}^{*} N_{\beta i}^{*} $ and $F^{}_{ji}\equiv \Delta m^2_{ji}L/(4E)$. For antineutrino oscillations $\overline{P}_{\alpha^{}\beta^{}}^{} \equiv P(\overline{\nu}_\alpha^{}\to\overline{\nu}_\beta^{})$ or the time-reversal processes $P_{\beta \alpha}^{} \equiv P(\nu_\beta^{} \to \nu_\alpha^{})$, the probabilities can be obtained by replacing $N\to N^*$ in Eq.~(\ref{eq:prob}). We emphasize that the starting point of the above derivations is the definition of flavor eigenstates in Eq.~(\ref{eq:flavor_state}), which are independent of the production and detection processes. 

As a direct consequence of the polar decomposition theorem, the non-unitary mixing matrix $N$ can be decomposed as $N=({\bf 1} - \eta) \cdot V'$~\cite{Broncano:2002rw,Fernandez-Martinez:2007iaa}, where $\eta$ is a Hermitian matrix and $V'$ is a unitary matrix. Another decomposition is $N = T \cdot V$~\cite{Xing:2007zj,Xing:2011ur}, where $V$ is a unitary matrix and $T$ is a lower-triangular matrix of the form
\begin{eqnarray}
	\label{eq:alpha_matrix}
	T \equiv {\bf 1} - \zeta = {\bf 1} - \begin{pmatrix}
		\zeta_{ee}^{} & 0 & 0 \\
		\zeta_{\mu e}^{} & \zeta_{\mu \mu}^{} & 0 \\
		\zeta_{\tau e}^{} &  \zeta_{\tau \mu}^{} & \zeta_{\tau\tau}^{} \\
	\end{pmatrix} \;.
\end{eqnarray}
Here the diagonal elements $\zeta_{\alpha\alpha}^{}$ (for $\alpha=e,\mu,\tau$) are real and positive, the off-diagonal elements $\zeta_{\alpha\beta}^{} \equiv \left|\zeta_{\alpha\beta}^{}\right| \exp(+{\rm i} \phi_{\alpha\beta}^{} )$ (for $\alpha\beta=\mu e, \tau e, \tau\mu$) are complex, and all remaining elements are zero. The unitary matrices $V$ and $V'$ in the lower-triangular and Hermitian parametrizations are not identical and contain different parameters. The relationship between these two parametrizations and their phenomenological implications are fully discussed in Refs.~\cite{Xing:2008fg,Luo:2008vp,Antusch:2009pm,Malinsky:2009df,Xing:2009ce,Xing:2012gd,Xing:2012kh,Li:2015oal,Blennow:2016jkn,Li:2018jgd,Huang:2019phr,Wang:2021rsi}. Although they are mathematically equivalent, the lower-triangular form is more suitable for discussing the non-unitarity effects on neutrino oscillations. Meanwhile, in the complete type-I seesaw model, physical parameters at the low-energy scale can be calculated with such a parametrization in a transparent way~\cite{Xing:2023kdj,Xing:2024xwb,Xing:2024gmy,Huang:2025ubs}. Therefore, we will choose the lower-triangular parametrization and specify the corresponding unitary matrix $V$ in the standard way advocated by Particle Data Group~\cite{ParticleDataGroup:2024cfk}:
\begin{eqnarray}
	V = \begin{pmatrix}
		c_{12}^{} c_{13}^{} & s_{12}^{} c_{13}^{} & s_{13}^{} {\rm e}^{-{\rm i} \widehat{\delta}_{\rm CP}^{}} \\
		-s_{12}^{} c_{23}^{}-c_{12}^{} s_{13}^{} s_{23}^{} {\rm e}^{{\rm i} \widehat{\delta}_{\rm CP}^{}} & +c_{12}^{} c_{23}^{}-s_{12}^{} s_{13}^{} s_{23}^{} {\rm e}^{{\rm i} \widehat{\delta}_{\rm CP}^{}} & c_{13}^{} s_{23}^{} \\
		+s_{12}^{} s_{23}^{}-c_{12}^{} s_{13}^{} c_{23}^{} {\rm e}^{{\rm i} \widehat{\delta}_{\rm CP}^{}} & -c_{12}^{} s_{23}^{}-s_{12}^{} s_{13}^{} c_{23}^{} {\rm e}^{{\rm i} \widehat{\delta}_{\rm CP}^{}} & c_{13}^{} c_{23}^{}
	\end{pmatrix} \cdot
	\begin{pmatrix}
		{\rm e}^{{\rm i} \widehat{\rho}} & 0 & 0 \\
		0 & {\rm e}^{{\rm i} \widehat{\sigma}} & 0 \\
		0 & 0 & 1
	\end{pmatrix} \;,
\end{eqnarray}
with $s_{ij}^{} \equiv \sin\widehat{\theta}_{ij}^{}$ and $c_{ij}^{} \equiv \cos\widehat{\theta}_{ij}^{}$ (for $ij=12,13,23$). However, the mixing angles $\{ \widehat{\theta}_{12}^{}, \widehat{\theta}_{13}^{}, \widehat{\theta}_{23}^{} \}$, the Dirac CP-violating phase $\widehat{\delta}_{\rm CP}^{}$ and two Majorana phases $\{\widehat{\rho},\widehat{\sigma}\}$ are distinct from those in the case of a unitary mixing matrix. The oscillation probabilities in Eq.~(\ref{eq:prob}) can be explicitly written out by using the following identities
\begin{eqnarray}
	\label{eq:TTdag}
	N N^\dagger = T T^\dagger = 
	\begin{pmatrix}
		\left( 1 - \zeta_{ee}^{} \right)^2 & - \left( 1 - \zeta_{ee}^{}\right)  \zeta_{\mu e}^{*} & - \left( 1 - \zeta_{ee}^{}\right)  \zeta_{\tau e}^{*} \vspace{0.1 cm} \\
		- \left( 1- \zeta_{ee}^{}\right)  \zeta_{\mu e}^{} & \left(1 - \zeta_{\mu \mu}^{}\right)^2 + \left|\zeta_{\mu e}^{}\right|^2 & - \left( 1 - \zeta_{\mu \mu}^{}\right)   \zeta_{\tau \mu}^{*} + \zeta_{\mu e}^{} \zeta_{\tau e}^{*} \vspace{0.1 cm} \\
		- \left( 1 - \zeta_{ee}^{}\right)  \zeta_{\tau e}^{} & - \left( 1 - \zeta_{\mu \mu}^{}\right)   \zeta_{\tau \mu}^{} + \zeta_{\mu e}^{*} \zeta_{\tau e}^{}  & \left(1 - \zeta_{\tau \tau}^{}\right)^2 + \left|\zeta_{\tau e}^{}\right|^2 + \left|\zeta_{\tau \mu}^{}\right|^2 
	\end{pmatrix} \;,
\end{eqnarray}
and
\begin{eqnarray}
	\label{eq:N}
	N = \begin{pmatrix}
		\left(1 - \zeta_{ee}^{} \right) V_{e1}^{} & \left(1-\zeta_{ee}^{} \right) V_{e2} & \left(1-\zeta_{ee}^{} \right) V_{e3} \vspace{0.25cm} \\ 
		\left(1 - \zeta_{\mu \mu}^{} \right)  V_{\mu 1}^{} - \zeta_{\mu e}^{} V_{e1}^{}  & \left(1-\zeta_{\mu \mu}^{} \right)  V_{\mu 2}^{} -\zeta_{\mu e}^{} V_{e2}^{}  &  \left(1-\zeta_{\mu \mu}^{} \right) V_{\mu 3}^{} -\zeta_{\mu e}^{} V_{e3}^{} \vspace{0.25cm} \\
		\begin{array}{l} \left(1 - \zeta_{\tau \tau}^{} \right)  V_{\tau 1}^{} \\ - \zeta_{\tau e}^{} V_{e1}^{} - \zeta_{\tau \mu}^{}  V_{\mu 1}^{}  \end{array} & \begin{array}{l}  \left(1-\zeta_{\tau \tau}^{} \right)  V_{\tau 2}^{} \\ -\zeta_{\tau e}^{} V_{e2}^{} - \zeta_{\tau \mu}^{}  V_{\mu 2}^{}  \end{array} & \begin{array}{l}  \left(1-\zeta_{\tau \tau}^{} \right)  V_{\tau 3}^{} \\ -\zeta_{\tau e}^{} V_{e3}^{} - \zeta_{\tau \mu}^{}  V_{\mu 3}^{}  \end{array} \\
	\end{pmatrix} \;.
\end{eqnarray}
These explicit expressions allow for a clear identification of which non-unitarity parameters appear in the oscillation probability of a given channel.

\subsection{Disappearance channels}

We first consider the disappearance channels $\nu_\alpha^{} \to \nu_\alpha^{}$. The oscillation probabilities can be easily read off from Eq.~(\ref{eq:prob}) as
\begin{eqnarray}
	P_{\alpha\alpha}^{} = \frac{1}{\left[\left(T T^\dagger\right)_{\alpha\alpha}^{}\right]^2} \left[\left|\left(T T^\dagger\right)_{\alpha\alpha}^{}\right|^2 - 4 \sum_{i<j} \left|N_{\alpha i}^{}\right|^2 \left|N_{\alpha j}^{}\right|^2 \sin^2 F_{ji}^{} \right] \;.
\end{eqnarray}
Here we have ${\cal N}_{\alpha\alpha}^{ij} = \left|N_{\alpha i}^{}\right|^2 \left|N_{\alpha j}^{}\right|^2$ and the relation $P_{\alpha\alpha}^{} = \overline{P}_{\alpha\alpha}^{}$ holds. From Eqs.~(\ref{eq:TTdag}) and (\ref{eq:N}), the probability for each flavor of (anti)neutrinos depends at most on the non-unitarity parameters in the corresponding row of the $\zeta$ matrix. 

For reactor antineutrino experiments, the disappearance channel $\overline{\nu}_e^{} \to \overline{\nu}_e^{}$ is relevant and the probability in the non-unitary case is
\begin{eqnarray}
	\label{eq:Pee}
	\overline{P}_{ee}^{} = P_{ee}^{} &=&  \frac{1}{\left(1 - \zeta_{ee}^{}\right)^4} \left[\left(1 - \zeta_{ee}^{}\right)^4 - 4 \left(1 - \zeta_{ee}^{}\right)^4 \sum_{i<j} \left|V_{e i}^{}\right|^2 \left|V_{e j}^{}\right|^2 \sin^2 F_{ji}^{} \right] \nonumber \\
	&=& 1 - 4 \sum_{i<j} \left|V_{e i}^{}\right|^2 \left|V_{e j}^{}\right|^2 \sin^2 F_{ji}^{} \;,
\end{eqnarray}
where we have made use of $N_{ei}^{} = \left(1 - \zeta_{ee}^{}\right) V_{ei}^{}$ and $\left(T T^\dagger\right)_{ee}^{} = \left(1 - \zeta_{ee}^{}\right)^2$. This is a very special result for the lower-triangular parametrization, since the probability is the same as that in the unitary case without involving any matrix elements of $\zeta$. Therefore, the measurements in the disappearance channel of $\nu_e^{}$ (or $\overline{\nu}_e^{}$) are insensitive to the non-unitarity effects, where the mixing parameters directly correspond to those in the unitary matrix $V$. On the other hand, in order to test and constrain the non-unitarity effects, it is essential to take into account the corresponding production and detection processes or other oscillation channels. 

For completeness, we give the expressions of oscillation probabilities $P_{\mu\mu}^{}$ and $P_{\tau \tau}^{}$. For muon (anti)neutrinos we have
\begin{eqnarray}
	P_{\mu \mu}^{}  = \overline{P}_{\mu\mu}^{} = \frac{1}{\left[\left(T T^\dagger\right)_{\mu \mu}^{}\right]^2} \left[\left|\left(T T^\dagger\right)_{\mu \mu}^{}\right|^2 - 4 \sum_{i<j} \left|N_{\mu i}^{}\right|^2 \left|N_{\mu j}^{}\right|^2 \sin^2 F_{ji}^{} \right] \;.
\end{eqnarray}
In this case, the non-unitarity parameters $\zeta_{\mu e}^{}$ and $\zeta_{\mu \mu}^{}$ appear. To see this clearly, we perform the series expansion of $P_{\mu \mu}^{}$ in terms of the elements of $\zeta$ that are regarded as small parameters:
\begin{eqnarray}
	P_{\mu \mu}^{} = \overline{P}_{\mu\mu}^{} &\approx& 1 - 4 \sum_{i<j} \left|V_{\mu i}^{}\right|^2 \left|V_{\mu j}^{}\right|^2 \sin^2 F_{ji}^{} \nonumber \\
	&& + 8 \sum_{i<j} {\rm Re} \left[\zeta_{\mu e}^{}  \left(V_{e i}^{} V_{\mu i}^* \left|V_{\mu j}^{}\right|^2 + V_{e j}^{}  V_{\mu j}^* \left|V_{\mu i}^{}\right|^2\right)  \right] \sin^2 F_{ji}^{} \;,
\end{eqnarray}
where the terms of ${\cal O}(\zeta^2)$ have been neglected. Similarly, the oscillation probabilities $P_{\tau\tau}^{}$ and $\overline{P}_{\tau \tau}^{}$ can be calculated as
\begin{eqnarray}
	P_{\tau \tau}^{}  = \overline{P}_{\tau \tau}^{} = \frac{1}{\left[\left(T T^\dagger\right)_{\tau \tau}^{}\right]^2} \left[\left|\left(T T^\dagger\right)_{\tau \tau}^{}\right|^2 - 4 \sum_{i<j} \left|N_{\tau i}^{}\right|^2 \left|N_{\tau j}^{}\right|^2 \sin^2 F_{ji}^{} \right] \;,
\end{eqnarray}
for which the approximate formulas read
\begin{eqnarray}
	P_{\tau \tau}^{} = \overline{P}_{\tau \tau}^{} &\approx& 1 - 4 \sum_{i<j} \left|V_{\tau i}^{}\right|^2 \left|V_{\tau j}^{}\right|^2 \sin^2 F_{ji}^{} \nonumber \\
	&& -8 \sum_{i<j} {\rm Re} \left[ \zeta_{\tau e}^{} \left(V_{e i}^{} V_{\tau i}^* \left|V_{\tau j}^{}\right|^2 + V_{e j}^{} V_{\tau j}^* \left|V_{\tau i}^{}\right|^2 \right) \right. \nonumber \\
	&& \left. + \zeta_{\tau \mu}^{}  \left(V_{\mu i}^{} V_{\tau i}^* \left|V_{\tau j}^{}\right|^2 + V_{\mu j}^{} V_{\tau j}^* \left|V_{\tau i}^{}\right|^2\right) \right] \sin^2 F_{ji}^{} \;, 
\end{eqnarray}
depending on  $\zeta_{\tau e}^{}$, $\zeta_{\tau \mu}^{}$ and $\zeta_{\tau \tau}^{}$ at the sub-leading order. 

\subsection{Appearance channels}

Appearance channels are of crucial importance for long-baseline accelerator neutrino experiments to discover leptonic CP violation, to which the measurements of reactor antineutrinos are in principle sensitive~\cite{Wang:2025ess}. In such circumstances, many more non-unitarity parameters in the matrix $\zeta$ are simultaneously involved. 

We start with the oscillation probability of $\nu_e^{} \to \nu_\mu^{}$ in the non-unitary case, which is given by
\begin{eqnarray}
\label{eq:Pemuexact}
	P_{e\mu}^{} &=& \frac{1}{\left(T T^\dagger\right)_{e e}^{} \left(T T^\dagger\right)_{\mu \mu}^{}} \left[\left|\left(T T^\dagger\right)_{e \mu}^{}\right|^2 - 4 \sum_{i<j} {\rm Re} \, {\cal N}_{e \mu}^{ij} \sin^2F_{ji}^{} + 2 \sum_{i<j} {\rm Im} \, {\cal N}_{e \mu}^{ij} \sin 2 F_{ji}^{} \right] \;, 
\end{eqnarray}
with ${\cal N}_{e\mu}^{ij} = N_{e i}^{} N_{\mu j}^{} N_{e j}^{*} N_{\mu i}^{*} $. Due to the following relations
\begin{eqnarray}
	{\cal N}_{e\mu}^{ij} = \left(T T^\dagger\right)_{ee}^{} V_{e i}^{} N_{\mu j}^{} V_{e j}^{*} N_{\mu i}^{*} \;, \quad \left|\left(T T^\dagger\right)_{e \mu}^{}\right|^2 = \left|T_{ee}^{} T_{\mu e}^* \right|^2 = \left(T T^\dagger\right)_{ee}^{} \left|T_{\mu e}^{}\right|^2 \;,
\end{eqnarray}
the factor $\left(T T^\dagger\right)_{ee}^{}$ in the oscillation probability is actually canceled out, and thus $\zeta_{ee}^{}$ is irrelevant. The probability for antineutrinos $\overline{P}_{e \mu}^{}$ is derived by simply changing the sign of the third term in the square brackets on the right-hand side of Eq.~(\ref{eq:Pemuexact}). Expanding it to the sub-leading order, we arrive at
\begin{eqnarray}
	\label{eq:Pemu}
	\overline{P}_{e\mu}^{} &\approx& -4 \sum_{i<j} {\rm Re} \left[V_{e i}^{} V_{\mu j}^{} V_{e j}^* V_{\mu i}^* - \left(\zeta_{\mu e}^{} V_{e i}^{} V_{\mu i}^* \left|V_{e j}^{}\right|^2 + \zeta_{\mu e}^* V_{e j}^* V_{\mu j}^{} \left|V_{e i}^{}\right|^2 \right) \right] \sin^2 F_{ji}^{}  \nonumber \\
	&& - 2 \sum_{i<j} {\rm Im} \left[V_{e i}^{} V_{\mu j}^{} V_{e j}^* V_{\mu i}^* - \left(\zeta_{\mu e}^{} V_{e i}^{} V_{\mu i}^* \left|V_{e j}^{}\right|^2 + \zeta_{\mu e}^* V_{e j}^* V_{\mu j}^{} \left|V_{e i}^{}\right|^2 \right) \right] \sin 2 F_{ji}^{} \;. 
\end{eqnarray}
The diagonal parameter $\zeta_{\mu \mu}^{}$ is not involved at ${\cal O}(\zeta)$, but will emerge at ${\cal O}(\zeta^2)$. 

The probability for $\overline{\nu}_e^{} \to \overline{\nu}_\tau^{}$ can be calculated in the same way and the approximate result is
\begin{eqnarray}
	\label{eq:Petau}
	\overline{P}_{e \tau}^{} &\approx& -4 \sum_{i<j} {\rm Re} \left[V_{e i}^{} V_{\tau j}^{} V_{e j}^* V_{\tau i}^* - \left(\zeta_{\tau e}^{} V_{e i}^{} V_{\tau i}^* \left|V_{e j}^{}\right|^2  + \zeta_{\tau e}^* V_{e j}^* V_{\tau j}^{} \left|V_{e i}^{}\right|^2  \right. \right. \nonumber \\
	&& \left. \left. \vphantom{\left|V_{e j}^{*}\right|^2} + \zeta_{\tau \mu}^{}  V_{e i}^{} V_{e j}^* V_{\mu j}^{} V_{\tau i}^* + \zeta_{\tau \mu}^* V_{e i}^{} V_{e j}^* V_{\mu i}^* V_{\tau j}^{}\right)\right] \sin^2 F_{ji}^{} \nonumber \\
	&& - 2 \sum_{i<j} {\rm Im} \left[V_{e i}^{} V_{\tau j}^{} V_{e j}^* V_{\tau i}^* - \left(\zeta_{\tau e}^{} V_{e i}^{} V_{\tau i}^* \left|V_{e j}^{}\right|^2  + \zeta_{\tau e}^* V_{e j}^* V_{\tau j}^{} \left|V_{e i}^{}\right|^2  \right. \right. \nonumber \\
	&& \left. \left. \vphantom{\left|V_{e j}^{*}\right|^2} + \zeta_{\tau \mu}^{}  V_{e i}^{} V_{e j}^* V_{\mu j}^{} V_{\tau i}^* + \zeta_{\tau \mu}^* V_{e i}^{} V_{e j}^* V_{\mu i}^* V_{\tau j}^{}\right)\right] \sin 2 F_{ji}^{} \;.
\end{eqnarray}
As one can observe from the above equation, the off-diagonal elements $\zeta_{\tau e}^{}$ and $\zeta_{\tau \mu}^{}$ appear in the expression at the sub-leading order. Finally, the probability for $\overline{\nu}_\mu^{} \to \overline{\nu}_\tau^{}$ is
\begin{eqnarray}
	\overline{P}_{\mu \tau}^{} &\approx& -4 \sum_{i<j} {\rm Re} \left[ \vphantom{\left(\left|V_{\mu j}^{}\right|^2 \right)} V_{\mu i}^{} V_{\tau j}^{} V_{\mu j}^* V_{\tau i}^* - \left(\zeta_{\mu e}^{} V_{e i}^{} V_{\mu j}^* V_{\tau i}^* V_{\tau j}^{} +\zeta_{\mu e}^* V_{e j}^* V_{\mu i}^{} V_{\tau i}^* V_{\tau j}^{} + \zeta_{\tau e}^{} V_{e j}^{} V_{\mu i}^{} V_{\mu j}^* V_{\tau i}^*  \right. \right. \nonumber \\
	&& \left. \left. + \zeta_{\tau e}^* V_{e i}^* V_{\mu i}^{} V_{\mu j}^* V_{\tau j}^{} + \zeta_{\tau \mu}^{}  V_{\mu i}^{} V_{\tau i}^* \left|V_{\mu j}^{}\right|^2  + \zeta_{\tau \mu}^* V_{\mu j}^* V_{\tau j}^{} \left|V_{\mu i}^{}\right|^2 \right) \right] \sin^2 F_{ji}^{} \nonumber \\
	&& - 2  \sum_{i<j} {\rm Im} \left[ \vphantom{\left(\left|V_{\mu j}^{}\right|^2 \right)} V_{\mu i}^{} V_{\tau j}^{} V_{\mu j}^* V_{\tau i}^* - \left(\zeta_{\mu e}^{} V_{e i}^{} V_{\mu j}^* V_{\tau i}^* V_{\tau j}^{} +\zeta_{\mu e}^* V_{e j}^* V_{\mu i}^{} V_{\tau i}^* V_{\tau j}^{} + \zeta_{\tau e}^{} V_{e j}^{} V_{\mu i}^{} V_{\mu j}^* V_{\tau i}^*  \right. \right. \nonumber \\
	&& \left. \left. + \zeta_{\tau e}^* V_{e i}^* V_{\mu i}^{} V_{\mu j}^* V_{\tau j}^{} + \zeta_{\tau \mu}^{}  V_{\mu i}^{} V_{\tau i}^* \left|V_{\mu j}^{}\right|^2  + \zeta_{\tau \mu}^* V_{\mu j}^* V_{\tau j}^{} \left|V_{\mu i}^{}\right|^2 \right) \right] \sin 2 F_{ji}^{} \;,
\end{eqnarray}
which depends on all three off-diagonal elements $\zeta_{\mu e}^{}$, $\zeta_{\tau e}^{}$ and $\zeta_{\tau \mu}^{}$. 

Now the oscillation probabilities in all oscillation channels are obtained. Before ending this section, we make some helpful remarks on them:
\begin{itemize}
	\item As they should be, all the oscillation probabilities are reduced to those in the unitary case by replacing the non-unitary mixing matrix $N$ with the unitary one $V$, the latter of which satisfies the unitarity condition $\left(V V^\dagger\right)_{\alpha\beta}^{} = \delta_{\alpha\beta}^{}$. However, in the full type-I seesaw model, one cannot simply take $T \to {\bf 1}$ or $\zeta \to {\bf 0}$, since this is only possible when the Yukawa couplings between right-handed neutrinos and the SM fields are switched off~\cite{Xing:2025bdm}. The limit $N \to V$ can be realized approximately in the type-I seesaw model by omitting the terms of ${\cal O}(\zeta)$ but retaining those of ${\cal O}(\sqrt{\zeta})$.
		
	\item When the oscillation length is set to zero, one has 
	\begin{eqnarray}
		P_{\alpha\beta}^{}(L=0) = \frac{\left|\left(T T^\dagger\right)_{\alpha\beta}^{}\right|^2}{\left(T T^\dagger\right)_{\alpha\alpha}^{} \left(T T^\dagger\right)_{\beta\beta}^{}} \;,
	\end{eqnarray}
	and $P_{\alpha\alpha}^{}(L=0) = 1$ holds even in the non-unitary case. However, for $\alpha\neq\beta$, the oscillation probabilities are still nonzero, for instance 
	\begin{eqnarray}
		P_{e \mu}^{}(L=0) = \frac{\left|\left(T T^\dagger\right)_{e\mu}^{}\right|^2}{\left(T T^\dagger\right)_{ee}^{} \left(T T^\dagger\right)_{\mu \mu}^{}} = \left|\zeta_{\mu e}^{}\right|^2 + {\cal O}(\zeta^3) \;.
	\end{eqnarray}
	This zero-distance effect implies that the near detectors for neutrino oscillation experiments are useful for constraining the non-unitarity parameters.

	\item The lower-triangular parametrization of the mixing matrix $N$ is also introduced in Ref.~\cite{Escrihuela:2015wra}, where the non-unitarity effects on neutrino oscillations are further discussed. Compared to the oscillation probabilities derived earlier, those calculated in Sec.~V of Ref.~\cite{Escrihuela:2015wra} lack the normalization factor $1/[(T T^\dagger)_{\alpha\alpha}^{} (T T^\dagger)_{\beta\beta}^{} ]$. After taking such a factor into account, the two results are consistent with each other.
\end{itemize}

In our discussions, we have separated the production, oscillation and detection of neutrinos into three independent parts. This is possible only when neutrino masses are much smaller than the characteristic energy scales of the production and detection processes. The oscillation probabilities can be calculated with the neutrino flavor eigenstates defined in Eq.~(\ref{eq:flavor_state}), which become independent of the production and detection. However, the non-unitarity effects on the latter must be included when evaluating the neutrino event rates and comparing them with experimental observations, which will be discussed in the following section.

\section{Event rates for reactor antineutrinos}

\label{sec:event}

\subsection{Cross sections and neutrino fluxes}

To examine the production and detection of neutrinos, we write down the Lagrangian for the CC and NC interactions of neutrinos with the non-unitary mixing matrix:
\begin{eqnarray}
	\label{eq:L_int}
	{\cal L}_{\rm CC}^{} = \frac{g}{\sqrt{2}} \overline{l_{\alpha}^{}} \gamma^\mu P_{\rm L}^{} N_{\alpha i}^{} \nu_i^{} W_\mu^{-} + {\rm h.c.} \;, \quad {\cal L}_{\rm NC}^{} = \frac{g}{2 \cos\theta_{\rm w}^{}} \overline{\nu_i^{}} \gamma^\mu P_{\rm L}^{} \left(N^\dagger N\right)_{ij}^{} \nu_j^{}  Z_\mu  \;,
\end{eqnarray}
where $g$ is the gauge coupling and $\theta_{\rm w}^{}$ is the weak mixing angle. The interaction terms are specified in the mass basis. However, if we consider neutrinos with specific flavors in the initial or final state, it is necessary to adopt the definition of flavor eigenstates given in Eq.~(\ref{eq:flavor_state}).

Let us first take the muon decay as an example, which contains two CC interaction vertices. In the low-energy limit, where the masses of weak gauge bosons are much higher than any other energy scales, the decay amplitude reads
\begin{eqnarray}
	{\cal M} \left( \mu^- \to e^- + \nu_{\mu}^{} + \overline{\nu}_e^{} \right) \simeq - \frac{g^2}{2 m_W^2} \left< e^-  \nu_{\mu}^{}  \overline{\nu}_e^{} \right| \left( \overline{\nu_i^{}} N_{\mu i}^* \gamma_\lambda^{} P_{\rm L}^{} \mu \right) \left( \overline{e} \gamma^\lambda P_{\rm L}^{} N_{e j}^{} \nu_j^{} \right) \left| \vphantom{\nu_\mu^{}} \mu^- \right> \;.
\end{eqnarray}
Then, with the help of the definition in Eq.~(\ref{eq:flavor_state}), the final states $\left< \nu_\mu^{} \right|$ and $\left< \overline{\nu}_e^{} \right|$ can be expressed in terms of mass eigenstates:
\begin{eqnarray}
	\left< \nu_\mu^{} \right| = \sum_m \left< \nu_m^{} \right| \frac{N_{\mu m}^{}}{\sqrt{\left(N N^\dagger\right)_{\mu \mu}^{}}} \;, \quad \left< \overline{\nu}_e^{} \right| = \sum_n \left< \nu_n^{} \right| \frac{N_{e n}^*}{\sqrt{\left(N N^\dagger\right)_{e e}^{}}} \;.
\end{eqnarray}
Making use of the orthogonality of mass eigenstates, we obtain
\begin{eqnarray}
	\label{eq:mu_deay_M}
	{\cal M} \left( \mu^- \to e^- + \nu_{\mu}^{} + \overline{\nu}_e^{} \right) = {\cal M}_{\rm SM}^{} \left( \mu^- \to e^- + \nu_{\mu}^{} + \overline{\nu}_e^{} \right) \sqrt{\left(N N^\dagger\right)_{e e}^{} \left(N N^\dagger\right)_{\mu \mu}^{}} 
\end{eqnarray}
after some straightforward calculations, where ${\cal M}_{\rm SM}^{}$ is the decay amplitude in the SM. Therefore, the Fermi coupling constant $G_\mu^{\rm exp}$ extracted from the measurements of the muon lifetime is no longer identical to the SM prediction $G_\mu^{\rm SM}$, but subject to the following modification~\cite{Antusch:2006vwa}
\begin{eqnarray}
	\label{eq:G_exp_G_SM}
	G_\mu^{\rm exp} = G_\mu^{\rm SM} \sqrt{\left(N N^\dagger\right)_{ee}^{} \left(N N^\dagger\right)_{\mu \mu}^{} } \;,
\end{eqnarray}
where $G_\mu^{\rm SM} = \left[g^2 / (4\sqrt{2} m_W^2)\right] \left(1 + \Delta r\right)$ at the one-loop level with $\Delta r$ containing all electroweak radiative corrections~\cite{Sirlin:1980nh}. If the measured Fermi coupling constant $G_\mu^{\rm exp}$ is adopted as an input parameter for the production and detection of neutrinos, the relation in Eq.~(\ref{eq:G_exp_G_SM}) must be taken into account to ensure a self-consistent analysis. Meanwhile, other parameters, such as the masses of charged leptons and the fine-structure constant, can be determined via the electromagnetic interactions, so they are not affected by the non-unitary mixing matrix.

The most important process for the detection of reactor antineutrinos is the inverse beta decay (IBD): $\overline{\nu}_e^{} + p \to e^+ + n$, which involves one CC interaction vertex of leptons. With the similar approach and using the relation in Eq.~(\ref{eq:G_exp_G_SM}), the cross section in the non-unitary case is $\sigma_{\rm IBD}^{} = \sigma_{\rm IBD}^{\rm SM} / \left(N N^\dagger\right)_{\mu \mu}^{} $. Here $\sigma_{\rm IBD}^{\rm SM}$ is the cross section in the SM evaluated with $G_\mu^{\rm exp}$ without unitarity violation. With the energy of the final-state positron $E_{e^+}^{} = E - m_n^{} + m_p^{}$ and its three-momentum $\left|{\bf p}_{e^+}^{}\right|=\sqrt{E_{e^+}^2-m_e^2}$, the IBD cross section in the SM is given by an empirical formula with a very high accuracy~\cite{Strumia:2003zx}:
\begin{eqnarray}
	\label{eq:sigma_IBD}
	\sigma_{\rm IBD}^{\rm SM} (E) \approx 10^{-43}~{\rm cm}^2 \left|{\bf p}_{e^+}^{}\right| E_{e^+}^{} E_{}^{-0.07056 + 0.02018 \ln E - 0.001953 \ln^3 E} \;,
\end{eqnarray}
with $G_\mu^{\rm exp} \approx 1.166 \times 10^{-5}~{\rm GeV}^{-2}$ being the input parameter and $E$ being the energy of the initial electron antineutrino. As implied by Eq.~(\ref{eq:TTdag}), the cross section $\sigma^{}_{\rm IBD}$ in the non-unitary case involves parameters $\left|\zeta_{\mu e}^{}\right|$ and $\zeta_{\mu \mu}^{}$. Therefore, they will be constrained in the reactor antineutrino experiments, such as JUNO and TAO.

Another process to detect antineutrinos is through the elastic scattering with electrons: $\overline{\nu}_\alpha^{} + e^- \to \overline{\nu}_\alpha^{} + e^-$ (for $\alpha=e, \mu, \tau$). In the SM, both CC and NC interactions are involved in the $\overline{\nu}_e^{}$-$e$ scattering and contribute to the cross section, while only NC interaction is available for $\overline{\nu}_\mu^{}$ and $\overline{\nu}_\tau^{}$. However, the situation with a non-unitary mixing matrix will be quite different. Take the elastic scattering $\overline{\nu}_{\mu}^{} + e^- \to \overline{\nu}_{\mu}^{} + e^-$ as an example. Following the same strategy for calculating the muon decay width and the IBD cross section, we get the NC amplitude
\begin{eqnarray}
	{\cal M}_{\rm NC}^{} \left(\overline{\nu}_{\mu}^{} + e^- \to \overline{\nu}_{\mu}^{} + e^-\right) &\simeq& - \frac{g^2}{4 m_W^2} \sum_{i,j} \left< \overline{\nu}_\mu^{} e^- \right| \left[\overline{\nu_i^{}} \gamma^\lambda P_{\rm L}^{} \left(N^\dagger N\right)_{ij}^{} \nu_j^{}\right] \left[\overline{e} \gamma_\lambda^{} \left(c_{\rm V}^{} - c_{\rm A}^{} \gamma^5 \right)  e \right] \left| \overline{\nu}_\mu^{} e^- \right> \nonumber \\
	&=& \frac{\left(N N^\dagger N N^\dagger\right)_{\mu\mu}^{}}{\left(N N^\dagger\right)_{\mu\mu}^{}}  {\cal M}_{\rm NC}^{\rm SM} \left(\overline{\nu} + e^- \to \overline{\nu} + e^-\right) \;,
\end{eqnarray}
with $c_{\rm V}^{} = -1/2 + 2 \sin^2\theta_{\rm w}^{}$ and $c_{\rm A}^{} = -1/2$ being the vector-type and the axial-vector-type coupling of electrons, respectively. Note that the definition of neutrino flavor states in Eq.~(\ref{eq:flavor_state}) is adopted, and ${\cal M}_{\rm NC}^{\rm SM} \left(\overline{\nu} + e^- \to \overline{\nu} + e^-\right)$ is the NC amplitude in the SM, which is identical for three types of antineutrinos. On the other hand, there is also a CC contribution
\begin{eqnarray}
	{\cal M}_{\rm CC}^{} \left(\overline{\nu}_{\mu}^{} + e^- \to \overline{\nu}_{\mu}^{} + e^-\right) &\simeq& - \frac{g^2}{2 m_W^2} \sum_{i,j} \left< \overline{\nu}_\mu^{} e^- \right| \left(\overline{\nu_i^{}} N_{e i}^* \gamma^\lambda P_{\rm L}^{} e\right) \left(\overline{e} \gamma_\lambda^{} P_{\rm L}^{} N_{e j}^{} \nu_j^{} \right) \left| \overline{\nu}_\mu^{} e^- \right> \nonumber \\
	&=& \frac{\left|\left(N N^\dagger\right)_{\mu e}^{}\right|^2}{\left(N N^\dagger\right)_{\mu\mu}^{}}  {\cal M}_{\rm CC}^{\rm SM} \left(\overline{\nu} + e^- \to \overline{\nu} + e^-\right) \;,
\end{eqnarray}
with ${\cal M}_{\rm CC}^{\rm SM} \left(\overline{\nu} + e^- \to \overline{\nu} + e^-\right)$ being the CC scattering amplitude in the SM. One may immediately realize that the CC contribution for $\overline{\nu}_\mu^{}$-$e$ scattering is at the next-to-leading order, since it is proportional to the off-diagonal element $\left(N N^\dagger\right)_{\mu e}^{}$. In the limit of $N \to V$, such a contribution vanishes because of $\left(V V^\dagger\right)_{\mu e}^{} = 0$. The discussion about the amplitude for $\overline{\nu}_\tau^{}$ is similar and the final result is
\begin{eqnarray}
	{\cal M}_{\rm NC}^{} \left(\overline{\nu}_{\tau}^{} + e^- \to \overline{\nu}_{\tau}^{} + e^-\right) &=& \frac{\left(N N^\dagger N N^\dagger\right)_{\tau \tau}^{}}{\left(N N^\dagger\right)_{\tau \tau}^{}}  {\cal M}_{\rm NC}^{\rm SM} \left(\overline{\nu} + e^- \to \overline{\nu} + e^-\right) \;, \nonumber \\
	{\cal M}_{\rm CC}^{} \left(\overline{\nu}_{\tau}^{} + e^- \to \overline{\nu}_{\tau}^{} + e^-\right) &=& \frac{\left|\left(N N^\dagger\right)_{\tau e}^{}\right|^2}{\left(N N^\dagger\right)_{\tau \tau}^{}}  {\cal M}_{\rm CC}^{\rm SM} \left(\overline{\nu} + e^- \to \overline{\nu} + e^-\right) \;.
\end{eqnarray}
However, for electron antineutrinos, both NC and CC interactions make the leading-order contributions. The scattering amplitudes in the non-unitary case can be found 
\begin{eqnarray}
	{\cal M}_{\rm NC}^{} \left(\overline{\nu}_{e}^{} + e^- \to \overline{\nu}_{e}^{} + e^-\right) &=& \frac{\left(N N^\dagger N N^\dagger\right)_{e e}^{}}{\left(N N^\dagger\right)_{e e}^{}}  {\cal M}_{\rm NC}^{\rm SM} \left(\overline{\nu} + e^- \to \overline{\nu} + e^-\right) \;, \nonumber \\
	{\cal M}_{\rm CC}^{} \left(\overline{\nu}_{e}^{} + e^- \to \overline{\nu}_{e}^{} + e^-\right) &=& \left(N N^\dagger\right)_{e e}^{} {\cal M}_{\rm CC}^{\rm SM} \left(\overline{\nu} + e^- \to \overline{\nu} + e^-\right) \;.
\end{eqnarray}

Adding the NC and CC amplitudes together, we are able to calculate the amplitude squared and derive the differential cross sections. In the SM, we have
\begin{eqnarray}
	\label{eq:nue_xscetion_SM}
	\frac{{\rm d} \sigma_{e}^{\rm SM}}{{\rm d} T_e^{}} &=& \frac{\left(G_{\mu}^{\rm exp}\right)^2 m_e^{}}{2 \pi} \left\{\frac{m_e^{} T_e^{}}{E^2} \left[\left(c_{\rm A}^{}+1\right)^2-\left(c_{\rm V}^{} +1\right)^2 \right] + \left( 1-\frac{T_e^{}}{E} \right) ^2 \left( c_{\rm A}^{} + c^{}_{\rm V} + 2 \right)^2 + \left( c_{\rm A}^{} - c^{}_{\rm V} \right)^2\right\}  \;, \nonumber \\
	\frac{{\rm d} \sigma_{\mu,\tau}^{\rm SM}}{{\rm d} T_e^{}} &=& \frac{\left(G_{\mu}^{\rm exp}\right)^2 m_e^{}}{2 \pi} \left\{\frac{m_e^{} T_e^{}}{E^{2}} \left( c_{\rm A}^2-c_{\rm V}^2\right) + \left(1-\frac{T_e^{}}{E}\right)^2 \left(c_{\rm A}^{} + c^{}_{\rm V}\right)^2 + \left( c_{\rm A}^{} - c^{}_{\rm V}\right)^2\right\} \;,
\end{eqnarray}
where $T^{}_e \equiv E^{}_e - m^{}_e$ stands for the recoil energy of the final-state electron, $G_\mu^{\rm exp}$ is the measured Fermi coupling constant, which is equal to $G_\mu^{\rm SM}$. In the case of a non-unitary flavor mixing, the cross sections can be written in terms of $G_\mu^{\rm SM}$ and other non-unitarity parameters. After considering the mapping relation in Eq.~(\ref{eq:G_exp_G_SM}), we get the cross sections of elastic $\overline{\nu}$-$e$ scattering in the non-unitary case as
\begin{eqnarray}
	\label{eq:nue_xsection}
	\frac{{\rm d} \sigma_{\beta}^{}}{{\rm d} T_e^{}} &=& \frac{\left(G_{\mu}^{\rm exp}\right)^2 m_e^{}}{2 \pi \left(N N^\dagger\right)_{ee}^{} \left(N N^\dagger\right)_{\mu \mu}^{}} \left\{ \vphantom{\left(\frac{T_e^{}}{E}\right)^2} \frac{m_e^{} T_e^{}}{E^{2}} \left[\left(\widetilde{c}_{\rm A}^\beta+ \widetilde{c}_{\rm C}^\beta \right)^2-\left(\widetilde{c}_{\rm V}^{\beta} + \widetilde{c}_{\rm C}^\beta \right)^2 \right] \right. \nonumber \\
	&&  \left. + \left(1-\frac{T_e^{}}{E}\right)^2 \left( \widetilde{c}_{\rm A}^{\beta} + \widetilde{c}_{\rm V}^{\beta} + 2 \widetilde{c}_{\rm C}^\beta \right)^2 + \left( \widetilde{c}_{\rm A}^{\beta} - \widetilde{c}_{\rm V}^{\beta} \right)^2\right\}
\end{eqnarray}
for $\beta = e , \mu, \tau$, together with
\begin{eqnarray}
	\widetilde{c}_{\rm V,A}^{\beta} \equiv \frac{\left(N N^\dagger N N^\dagger\right)_{\beta \beta}^{}}{\left(N N^\dagger\right)_{\beta \beta}^{}} \times c_{\rm V,A}^{} \;, \quad \widetilde{c}_{\rm C}^\beta \equiv \frac{\left|\left(N N^\dagger\right)_{\beta e}^{}\right|^2}{\left(N N^\dagger\right)_{\beta \beta}^{}}
\end{eqnarray}
being the modified coupling constants of the vector-type, axial-vector-type, and CC-type. In the unitary case, we have $\widetilde{c}_{\rm V,A}^{\beta} \to c_{\rm V,A}^{}$ and $\widetilde{c}_{\rm C}^\beta \to \delta_{\beta e}^{}$, and the cross sections reduce to those in the SM as given in Eq.~(\ref{eq:nue_xscetion_SM}).

Expanding the coefficients $\widetilde{c}_{\rm V,A,C}^{\beta}$ with respect to the small elements of $\zeta$, we arrive at
\begin{eqnarray}
	\widetilde{c}_{\rm V,A}^{\beta} \approx \left(1 - 2 \zeta_{\beta \beta}^{} \right) c_{\rm V,A}^{} \;, \quad \widetilde{c}_{\rm C}^\beta \approx \begin{cases}
		\left|\zeta_{\beta e}^{}\right|^2 & \beta=\mu,\tau \\ 
		1 - 2 \zeta_{e e}^{} & \beta = e
	\end{cases} \;.
\end{eqnarray}
Due to the stringent experimental constraints on the non-unitarity parameters in the lower-triangular matrix $\zeta$~\cite{Blennow:2016jkn,Blennow:2023mqx}, as well as the observed relation $\widetilde{c}_{\rm V,A}^{\mu,\tau} \gg \widetilde{c}_{\rm C}^{\mu,\tau}$, it is reasonable to further expand the cross sections in Eq.~(\ref{eq:nue_xsection}) as
\begin{eqnarray}
	\label{eq:ESxsection}
	\frac{{\rm d} \sigma_{\beta}^{}}{{\rm d} T_e^{}} \approx \frac{{\rm d} \sigma_{\beta}^{\rm SM}}{{\rm d} T_e^{}} \times \begin{cases}
		1 - 2 \left(\zeta_{e e}^{} - \zeta_{\mu \mu}^{}\right) & \beta = e \\
		1 + 2 \left(\zeta_{e e}^{} - \zeta_{\mu \mu}^{}\right) & \beta = \mu \\
		1 + 2 \left( \zeta_{e e}^{} + \zeta_{\mu \mu}^{} - 2 \zeta_{\tau \tau}^{} \right) & \beta = \tau
	\end{cases} \;,
\end{eqnarray}
or equivalently 
\begin{eqnarray}
	\label{eq:ES}
	\frac{{\rm d} \sigma_{\beta}^{}}{{\rm d} T_e^{}} \approx \frac{{\rm d} \sigma_{\beta}^{\rm SM}}{{\rm d} T_e^{}} \times \frac{\left[\left(N N^\dagger\right)_{\beta \beta}^{}\right]^2}{\left(N N^\dagger\right)_{ee}^{} \left(N N^\dagger\right)_{\mu \mu}^{}}  \;.
\end{eqnarray}
In this way, one can see clearly the dependence of the cross sections on the non-unitarity parameters, and these results can be used for numerical calculations with a very good precision. 

In the end, let us investigate how non-unitarity effects modify the flux of electron antineutrinos from nuclear reactors. In practice, the antineutrino flux is determined by the output of thermal power. Given the amount of thermal energy released in each fission of four main isotopes, i.e., ${}^{235} {\rm U}$, ${}^{238} {\rm U}$, ${}^{239} {\rm Pu}$ and ${}^{241} {\rm Pu}$, the total number of the fissions can be figured out, and thus the antineutrino flux is obtained. Therefore, the antineutrino flux ${\rm d} \Phi_{\overline{\nu}_e^{}}^{} / {\rm d} E$ is independent of whether the lepton flavor mixing matrix is unitary or not, and thus does not need any corrections~\cite{Aloni:2022ebm}.\footnote{This conclusion is different from the statements in previous works, such as those in Refs.~\cite{Antusch:2006vwa,Dutta:2019hmb}, where a normalization factor $\left(N N^\dagger\right)_{ee}^{}$ is introduced to correct the neutrino flux in the so-called ``SM" case. In Ref.~\cite{CentellesChulia:2024sff}, the detection of accelerator neutrinos by the NC $\nu$-$e$ scattering is discussed, where the non-unitarity effects on the production, oscillation and detection of neutrinos are all included. As a result, correction factors will also contribute to the CC production process, which are absent in our circumstances.} For this reason, the flux of the electron antineutrinos from the reactor core $i$ can be calculated as
\begin{eqnarray}
	\label{eq:nu_flux}
	\frac{{\rm d} \Phi_{\overline{\nu}_e^{}}^i}{{\rm d} E_{\overline{\nu}_e^{}}^{}} = \frac{P_i^{}}{\displaystyle \sum_j f_j^{} \epsilon_j^{}} \sum_j f_j^{} S_j^{} (E_{\overline{\nu}_e^{}}^{}) \;,
\end{eqnarray}
where $P_i^{}$ is the thermal power for each reactor core, $\{f_j^{}, \epsilon_j^{},  S_j^{}(E_{\overline{\nu}_e^{}}^{})\}$ stand respectively for the fission fraction, the thermal energy released in each fission, and the neutrino energy spectrum per fission for the $j$-th isotope. For numerical calculations, we adopt $f_j^{} = \{0.561, 0.076, 0.307, 0.056\}$ and $\epsilon_j^{} = \{202.36, 205.99, 211.12, 214.26\}~{\rm MeV}$~\cite{JUNO:2015zny,JUNO:2020ijm}. Meanwhile, the energy spectrum is given by a 5th-order polynomial parametrization, and the values of coefficients $\alpha_{p j}^{}$ for the isotope $j$ at the $(p-1)$-th order can be found in Table VI of Ref.~\cite{Mueller:2011nm}. 

\subsection{IBD events at TAO}

TAO is located close to the Taishan nuclear power plants (NPPs), which have two 4.6~GW reactor cores. The central detector of TAO contains 2.8 tons of liquid scintillator (LS) with the number of free protons $N_p^{} \approx 2 \times 10^{29}$, where a 12\% hydrogen mass fraction is assumed. Its baselines to two cores are 44~m and 217~m, respectively. It is designed to reach the energy resolution as high as 1.5\% at 1~MeV. Therefore, the IBD event rate at TAO is
\begin{eqnarray}
	\frac{{\rm d} N_{\rm TAO}^{}}{{\rm d} E_{\rm obs}^{}} = N_{p}^{} t \sum_i \int_{E_{\rm thr}^{}}^\infty {\rm d} E \, \frac{{\rm d} \Phi_{\overline{\nu}_e^{}}^{i}}{{\rm d} E} \frac{\overline{P}_{e e}^{}(E,L)}{4 \pi L_i^2} \sigma_{\rm IBD}^{}(E) {\cal G}\left(E_{\rm obs}^{}; E_{\rm vis}^{},\delta_{\rm E}^{\rm TAO}\right) \;,
\end{eqnarray}
where $t$ is the data-taking time, $E_{\rm thr}^{} \approx 1.8~{\rm MeV}$ is the energy threshold for the IBD process, and the response function ${\cal G}\left(E_{\rm obs}^{}; E_{\rm vis}^{},\delta_{\rm E}^{\rm TAO}\right)$ is taken as the Gaussian function of the observed energy $E_{\rm obs}^{}$ with the visible energy $E_{\rm vis}^{} \equiv E_e^{} + m_e^{} = E - m_n^{} + m_p^{} + m_e^{}$ and the energy resolution of TAO detector $\delta_{\rm E}^{\rm TAO}$ being the expectation value and the standard derivation, respectively. Finally, one should sum up the contributions from all nuclear cores $i$ with the antineutrino flux ${\rm d} \Phi_{\overline{\nu}_e^{}}^{i} / {\rm d} E$ and the distance to the detector $L_i^{}$. In the non-unitary case, the cross section of IBD is modified, while the oscillation probability and the antineutrino fluxes keep unchanged. Inserting the matrix elements of $TT^\dagger$ into Eq.~(\ref{eq:TTdag}), we arrive at
\begin{eqnarray}
	\label{eq:nurate_TAO}
	\frac{{\rm d} N_{\rm TAO}^{}}{{\rm d} E_{\rm obs}^{}} = \frac{N_{p}^{} t}{\left[\left|\zeta_{\mu e}^{}\right|^2 + \left(1 - \zeta_{\mu \mu}^{}\right)^2\right]} \sum_i \int_{E_{\rm thr}^{}}^\infty {\rm d} E \, \frac{{\rm d} \Phi_{\overline{\nu}_e^{}}^{i}}{{\rm d} E} \frac{\overline{P}_{e e}^{}(E,L)}{4 \pi L_i^2} \sigma_{\rm IBD}^{\rm SM}(E) {\cal G} \left( E_{\rm obs}^{}; E_{\rm vis}^{},\delta_{\rm E}^{\rm TAO}\right)  \;. \quad
\end{eqnarray}
So one can place constraints on $\left|\zeta_{\mu e}^{}\right|$ and $\zeta_{\mu \mu}^{}$ through the measurements of antineutrinos in TAO. With the oscillation probability in Eq.~(\ref{eq:Pee}), the cross section in the SM in Eq.~(\ref{eq:sigma_IBD}), and the antineutrino flux in Eq.~(\ref{eq:nu_flux}), we can evaluate the event spectrum as a function of $\left|\zeta_{\mu e}^{}\right|$ and $\zeta_{\mu \mu}^{}$. 

To analyze the sensitivity of TAO to those relevant non-unitarity parameters, we construct a simplified $\chi^2$ function as
\begin{eqnarray}
	\chi_{\rm TAO}^2 = \sum_i \frac{\left(N_{\rm TAO}^{i, {\rm th}} - N_{\rm TAO}^{i, {\rm exp}}\right)^2}{N_{\rm TAO}^{i, {\rm exp}}} \;,
\end{eqnarray}
where $i$ runs over the numbers of energy bins, $N_{\rm TAO}^{i, {\rm exp}}$ is the number of observed events in each bin, and $N_{\rm TAO}^{i, {\rm th}}$ is the theoretically expected number as the function of non-unitarity parameters $\left|\zeta_{\mu e}^{}\right|$ and $\zeta_{\mu \mu}^{}$. The statistical uncertainty in each bin is set as $\sqrt{N_{\rm TAO}^{i, {\rm exp}}}$ and the systematic uncertainties are omitted for simplicity. In addition, the values of oscillation parameters in the unitary matrix $V$, i.e., three mixing angles and one Dirac CP-violating phase, together with two mass-squared differences, are fixed. We take the latest results of $\sin^2 \widehat{\theta}_{12}^{} = 0.3092$ and $\Delta m_{21}^2 = 7.50 \times 10^{-5}~{\rm eV}^2$ from JUNO~\cite{JUNO:2025gmd}, while others are quoted from the global analysis of current neutrino oscillation data~\cite{Esteban:2024eli,Capozzi:2025wyn}, namely,
\begin{eqnarray}
	\sin^2 \widehat{\theta}_{13}^{} = 0.02215 \;, \quad \sin^2 \widehat{\theta}_{23}^{} = 0.470 \;, \quad \widehat{\delta}_{\rm CP}^{} = 212^\circ \;, \quad \Delta m_{31}^2 = + 2.513 \times 10^{-3}~{\rm eV}^2 \;,
\end{eqnarray}
in the case of normal mass ordering (NO). The experimental data are generated with the above chosen parameters and zero values of non-unitarity parameters. One may notice from Eq.~(\ref{eq:nurate_TAO}) that the dependence on non-unitarity parameters is quite simple, which can be regarded as an overall correction to the total number of events without any approximation. With the date-taking time $t$ and about $N_{\rm TAO}^{} \approx 6300$ events per day in TAO, we obtain
\begin{eqnarray}
	\label{eq:chi2_TAO_numerical}
	\left|\zeta_{\mu e}^{}\right|^2 + \left(1 - \zeta_{\mu \mu}^{}\right)^2 = \frac{1}{1 \pm \sqrt{ \chi_{\rm TAO}^2 / (N_{\rm TAO}^{} t)}}\simeq 1 \pm 0.0126 \sqrt{\chi_{\rm TAO}^2/t} \;.
\end{eqnarray} 
Thus, the longer the experiment runs, the stronger the constraints on parameters will be. Compared to JUNO, for which the rate is 60 IBD events per day, TAO will limit the parameter space and provide us with much stricter upper bounds.

\begin{figure}[t]
	\centering
	\includegraphics[scale=0.67]{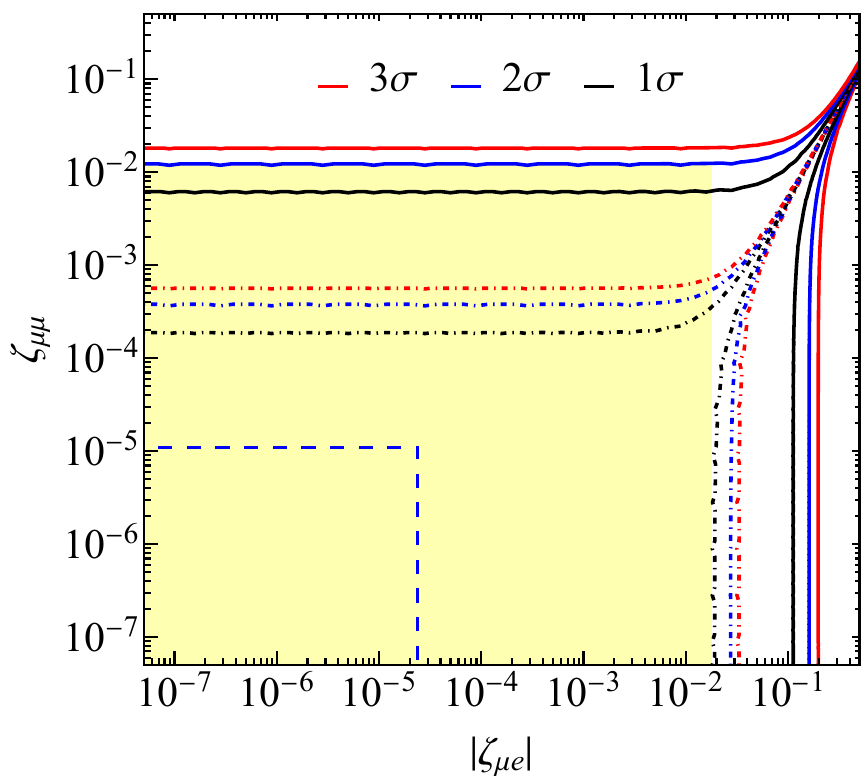}
	\vspace{-0.2cm}
	\caption{The sensitivity of TAO to the non-unitarity parameters $\left|\zeta_{\mu e}^{}\right|$ and $\zeta_{\mu \mu}^{}$ in the case of normal neutrino mass ordering. The black, blue and red solid (dot-dashed) curves represent the $1\sigma$, $2\sigma$ and $3\sigma$ contours with one-day (three-year) data, respectively. The yellow shaded region indicates constraints $\left|\zeta_{\mu e}^{}\right| < 1.8 \times 10^{-2}$ and $\zeta_{\mu \mu}^{} < 1.2 \times 10^{-2}$ from previous neutrino oscillation experiments at the 90\% confidence level ~\cite{Blennow:2025qgd}. As a comparison, the area enclosed by the blue dashed lines denotes the region of $\left|\zeta_{\mu e}^{}\right| < 2.4 \times 10^{-5}$ and $\zeta_{\mu \mu}^{} < 1.1 \times 10^{-5}$, which are the $2\sigma$ upper limits obtained from the global-fit analysis of current flavor and electroweak precision observables~\cite{Blennow:2023mqx}. }
	\label{fig:tao}
\end{figure}

In Fig.~\ref{fig:tao} we show the contours in the plane of $\left\{\left|\zeta_{\mu e}^{}\right|, \zeta_{\mu \mu}^{}\right\}$ at $1\sigma$, $2\sigma$ and $3\sigma$ confidence level (CL) as black, blue and red solid (dot-dashed) curves, respectively, with TAO's data in one day (three years). Some helpful comments on the numerical results are in order.
\begin{itemize}
	\item If $\left| \zeta_{\mu e}^{}\right|$ and $\zeta_{\mu \mu}^{}$ are both small enough, it is the diagonal element $\zeta_{\mu \mu}^{}$ that dominates the contribution to $\chi_{\rm TAO}^2$, since it appears as a linear term in the left-hand side of Eq.~(\ref{eq:chi2_TAO_numerical}), while $\left|\zeta_{\mu e}^{}\right|$ enters as a quadratic term. However, when $\left|\zeta_{\mu e}^{}\right|$ is relatively larger and close to 1, $\zeta_{\mu \mu}^{}$ should also be sufficiently large, and thus the cancellation between these two terms ensures that the right-hand side of Eq.~(\ref{eq:chi2_TAO_numerical}) remains unchanged. This observation explains the shapes of the contours.
	
	\item The constraints on two parameters from previous neutrino oscillation experiments are shown as the yellow region for comparison, with $\left|\zeta_{\mu e}^{}\right| < 1.8 \times 10^{-2}$ and $\zeta_{\mu \mu}^{} < 1.2 \times 10^{-2}$ at 90\% CL~\cite{Blennow:2025qgd}. It can be observed that in the region where $\left|\zeta_{\mu e}^{}\right|$ is very small and $\zeta_{\mu \mu}^{} \lesssim 10^{-2}$, TAO itself have already provided a slightly tighter constraint with one day of data. They will be improved by about one order of magnitude after three years of data-taking, much better than the constraints from current oscillation experiments. We also plot the upper limits from the global analysis of current flavor and electroweak precision observables by the blue dashed line. In the neutrino mass model with three heavy neutrinos, the upper bounds in the NO case are given as $\left|\zeta_{\mu e}^{}\right| < 2.4 \times 10^{-5}$ and $\zeta_{\mu \mu}^{} < 1.1 \times 10^{-5}$ at the $2\sigma$ CL~\cite{Blennow:2023mqx}. These are much stronger limits than those derived from neutrino oscillation experiments, demonstrating the necessity of performing the global analysis of electroweak precision measurements and higher-order calculations in the full theory, as in the SM~\cite{Marciano:1980pb,Sarantakos:1982bp,Strumia:2003zx,Ricciardi:2022pru,Huang:2024rfb}.
\end{itemize}

\subsection{JUNO}

The JUNO detector is located at $L_{\rm JUNO}^{} \approx 52.5~{\rm km}$ away from the Taishan and Yangjiang NPPs, with two 4.6~GW and six 2.9~GW cores, respectively. The central detector contains 20 kiloton LS with the total electron number $N_e^{} \approx 6.73 \times 10^{33}$, and its energy resolution is about 3\% at 1~MeV. Although JUNO can also detect electron antineutrinos through the IBD process, the number of events per day is smaller by two orders of magnitude than that at TAO. Meanwhile, there are only about 10 events per day for elastic scattering with electrons, even fewer than IBD events, so it is not an ideal way to constrain non-unitarity effects by detecting $\overline{\nu}_e^{}$ at JUNO. On the other hand, thanks to the long baseline, it is possible to detect $\overline{\nu}_\mu^{}$ and $\overline{\nu}_\tau^{}$ from appearance channels by the elastic scattering with electrons. In this case, there will be more non-unitarity parameters involved. The event rate of the elastic $\overline{\nu}_\beta^{}$-$e$ scattering (for $\beta = \mu,\tau$) can be evaluated through 
\begin{eqnarray} 
		\frac{{\rm d} N_{\rm JUNO}^{\beta}}{{\rm d} E_{\rm obs}^{}} = \frac{N_e^{} t}{4\pi L_{\rm JUNO}^2} \sum_i \int_0^{\infty} {\rm d} T_e^{} \, {\cal G} \left(E_{\rm obs}^{}; T_e^{}, \delta_{\rm E}^{\rm JUNO}\right) \int_{E_{\rm min}^{}}^{\infty} {\rm d} E \, \frac{{\rm d} \Phi_{\overline{\nu}_e^{}}^{i} }{{\rm d} E} \overline{P}_{e \beta}^{} \left(E, L_{\rm JUNO}^{}\right)  \frac{{\rm d} \sigma_{\beta}^{}}{{\rm d} T_e^{}} \;, \quad
\end{eqnarray}
where $E_{\rm min}^{} = T_e^{}/2 + \sqrt{T_e^{} \left(2m_e^{} + T_e^{}\right)} /2$ is the minimal neutrino energy for a given recoil energy $T_e^{}$ of the final-state electron, and the notations of other quantities are the same as for TAO. With the oscillation probabilities derived in Eqs.~(\ref{eq:Pemu}) and (\ref{eq:Petau}), the differential cross sections expressed in Eq.~(\ref{eq:ESxsection}) and the same numerical setup as for TAO, we construct the $\chi_{\rm JUNO}^2$ function in the same way 
\begin{eqnarray}
	\chi_{\rm JUNO}^2 = \sum_i \frac{\left(N_{\rm JUNO}^{i, {\rm th}} - N_{\rm JUNO}^{i,{\rm exp}} \right)^2}{N_{\rm JUNO}^{i,{\rm exp}} } \;,
\end{eqnarray}
where $N_{\rm JUNO}^{i, {\rm th}} $ is the total number of predicted events of the elastic $\overline{\nu}_\mu^{}$-$e$ and $\overline{\nu}_\tau^{}$-$e$ scattering in each energy bin as the function of all relevant non-unitarity parameters, while $N_{\rm JUNO}^{i,{\rm exp}}$ is the measured event numbers. The systematic uncertainty is also omitted and the statistical uncertainty is $\sqrt{N_{\rm JUNO}^{i,{\rm exp}}}$. To illustrate the capability for JUNO of constraining those parameters, we shall only discuss the following two cases.
\begin{itemize}
	\item {\bf Case I}. We retain all diagonal elements $\{\zeta_{ee}^{}, \zeta_{\mu \mu}^{}, \zeta_{\tau \tau}^{}\}$ and set others to zero. Since oscillation probabilities for appearance channels in Eqs.~(\ref{eq:Pemu}) and (\ref{eq:Petau}) only depend on off-diagonal elements up to ${\cal O}(\zeta^2)$, they will not be significantly modified in the non-unitary case, while the cross sections in Eq.~(\ref{eq:ESxsection}) are accordingly corrected. As an overall factor, the observed event rates can be evaluated as
	\begin{eqnarray}
		\frac{{\rm d} N_{\rm JUNO}^{\mu}}{{\rm d} E_{\rm obs}^{}} &\simeq& \left[1 + 2 \left(\zeta_{e e}^{} - \zeta_{\mu \mu}^{} \right) \right] \frac{{\rm d} N_{\rm JUNO}^{\rm \mu, SM}}{{\rm d} E_{\rm obs}^{}} \;, \nonumber \\
		\frac{{\rm d} N_{\rm JUNO}^{\tau}}{{\rm d} E_{\rm obs}^{}} &\simeq& \left[1 + 2 \left(\zeta_{e e}^{} + \zeta_{\mu \mu}^{} - 2\zeta_{\tau \tau}^{} \right) \right] \frac{{\rm d} N_{\rm JUNO}^{\rm \tau, SM}}{{\rm d} E_{\rm obs}^{}} \;.
	\end{eqnarray}
	Since the SM cross sections of $\overline{\nu}_\mu^{}$ and $\overline{\nu}_\tau^{}$ are identical at the leading order, and the oscillation probabilities from the reactor to JUNO detector differ marginally, the $\chi_{\rm JUNO}^2$ function can be roughly estimated as
	\begin{eqnarray}
		\chi_{\rm JUNO}^2 \simeq 8 t N_{\rm JUNO}^{\rm \mu,SM} \left(\zeta_{e e}^{} - \zeta_{\tau \tau}^{}\right)^2 \;,
	\end{eqnarray}
	in which the event numbers are $N_{\rm JUNO}^{\rm \mu,SM} \approx N_{\rm JUNO}^{\rm \tau,SM} \approx 2 $ per day. After three years, this leads to an upper bound $\left|\zeta_{e e}^{} - \zeta_{\tau \tau}^{}\right| \lesssim 0.76~(2.3) \times 10^{-2}$ at the $1\sigma~(3\sigma)$ CL.
	
	\item {\bf Case II}. We focus on the same non-unitarity parameters relevant for TAO and thus retain only $|\zeta_{\mu e}^{}|$ and $\zeta_{\mu \mu}^{}$, while setting others to zero. By noticing the minus sign in front of $\zeta_{\mu \mu}^{}$ in Eq.~(\ref{eq:ESxsection}) along with the identical cross sections of $\overline{\nu}_\mu^{}$ and $\overline{\nu}_\tau^{}$ in the SM, the $\chi_{\rm JUNO}^2$ function is independent of $\zeta_{\mu \mu}^{}$ and can be approximated as 
	\begin{eqnarray}
		\chi_{\rm JUNO}^2 \simeq \frac{1}{2} |\zeta_{\mu e}^{}|^2 t N_{\rm JUNO}^{\rm \mu,SM} \;.
	\end{eqnarray}
	Meanwhile, it is also proportional to the product of the matrix elements of $V$ due to the oscillation probability in Eq.~(\ref{eq:Pemu}), which is of ${\cal O}(1)$. As a result, the upper limit at $1\sigma~(3\sigma)$ CL after three years is estimated as $|\zeta_{\mu e}^{}| \lesssim 3.0~(9.1) \times 10^{-2}$.
\end{itemize}

Although the constraining power of JUNO on the non-unitarity parameters is limited by the total statistics, our investigation offers an affirmative answer to whether reactor antineutrino experiments are sensitive to unitarity violation. In particular, we stress that it is necessary to reconsider the production and detection of neutrinos when the flavor mixing matrix is non-unitary.

\subsection{Further discussions}

For the reactor antineutrino experiments measuring the IBD event rates in both near and far detectors and making use of the event ratio, such as the Daya Bay experiment, the $\chi^2$ function is basically constructed as follows~\cite{DayaBay:2012fng}
\begin{eqnarray}
	\chi^2 = \sum_i \frac{\left(N_{i}^{\rm FD} - \omega_i^{} N_{i}^{\rm ND}\right)^2}{N_{i}^{\rm FD}} \;,
\end{eqnarray}
where $N_i^{\rm FD}$ is the measured IBD events in each energy bin at the far detector and $N_i^{\rm ND}$ is that in the near detector. The weight factor $\omega_i^{} \equiv \overline{P}_{e e}^{i,{\rm FD}} (E_i^{}, L_{\rm FD}^{}) / \overline{P}_{e e}^{i, {\rm ND}} (E_i^{}, L_{\rm ND}^{})$, which is essentially the oscillation probability, is utilized to predict the event rate at the far detector. In reality, in addition to statistical uncertainties, systematic uncertainties and various types of backgrounds need to be taken into account. Since the oscillation probability $\overline{P}_{e e}^{}$ does not contain any parameter describing the non-unitarity effect, the total measurements are independent of whether the flavor mixing matrix is unitary or not. Such an approach is beneficial for the precise measurement of the mixing angles. On the other hand, in order to enhance the sensitivity to non-unitarity effects, we can individually analyze the event rates at the far (JUNO) or near (TAO) detector as what we have done in previous sections. The dependence on non-unitarity parameters in this case is quite pronounced, where the vast majority of parameters enter into the calculations of event rates. If a more detailed analysis is to be conducted, it should also incorporate the statistical effects arising from uncertainties in reactor antineutrino fluxes, as well as other potential influencing factors.

As the end of this section, we briefly compare the different strategies for detecting neutrinos via NC interactions in the literature. The most apparent distinction lies in the fact that the NC interaction could not distinguish among the three flavors of neutrinos or antineutrinos. Consequently, the oscillation probability $P_{\nu_\alpha^{} \nu_i^{}}^{}$ in Ref.~\cite{Antusch:2006vwa} or $P_{\nu_\alpha^{}\to \nu_{\rm NC}^{}}^{} $ in Ref.~\cite{Blennow:2025qgd} (derived from Ref.~\cite{Dutta:2019hmb}) are computed by summing over the final mass eigenstates $\nu_i^{}$, and the cross section is directly provided in the mass states 
\begin{eqnarray}
	\sigma_i^{\rm NC} = \sigma_{\rm NC}^{\rm SM} \sum_j \left|\left(N^\dagger N\right)_{ij}^{}\right|^2 \;,
\end{eqnarray}
where $\sigma_{\rm NC}^{\rm SM}$ is the NC cross section in the SM. In Ref.~\cite{Antusch:2006vwa} the oscillation probability is defined as $\widehat{P}_{\nu_\alpha^{} \nu_i^{}}^{} (E,L) \equiv \left|N_{\alpha i}^{}\right|^2$ after amputating the normalization factor in the denominator. This implies the produced neutrinos propagate as mass eigenstates, and are already decoherent by the time they reach the detector without oscillations. As a result, they individually participate in the NC detection process, such that a final summation over the mass eigenstates $\nu_i^{}$ gives rise to the final event rate. On the other hand, in Refs.~\cite{Dutta:2019hmb,Blennow:2025qgd}, the amputated probabilities are defined as 
\begin{eqnarray}
	\widehat{P}_{\nu_\alpha^{}\to\nu_{\rm NC}^{}}^{} (E,L) = \sum_j \left|\sum_i N_{\alpha i}^* {\rm e}^{-{\rm i}E_i^{} L} \left(N^\dagger N\right)_{ij}^{}\right|^2 \;.
\end{eqnarray}
This assumes that when neutrinos propagate to the detector as mass eigenstates, they undergo NC interactions with the targets in a coherent manner, and the total probability of detection via NC is obtained through the summation over the final mass eigenstates.

However, given a baseline length and the neutrino beam energy, the probability of oscillating to a specific flavor eigenstate should be determined as in Eq.~(\ref{eq:prob}). Meanwhile, the amplitudes and cross sections for NC interactions with the specific flavor eigenstate can also be directly computed and remain unaffected by whether each flavor is distinguishable, as long as the definition of neutrino flavor states in Eq.~(\ref{eq:flavor_state}) is taken. Instead of treating the oscillation and the detection as a whole, we separate the oscillation and the detection into two parts, and the calculation for each part in the non-unitary case can be handled independently. This approach is in accordance with the usual treatment for a unitary mixing matrix, and we believe the interpretation presented here is both physically transparent and logically self-consistent.

	\section{Summary}
	
	\label{sec:sum}
	
	In this paper, we investigate how to probe the non-unitarity of lepton flavor mixing matrix in the reactor antineutrino experiments, such as JUNO and TAO. Starting with the definition of neutrino flavor states in Eq.~(\ref{eq:flavor_state}), we derive the oscillation probabilities induced by a non-unitary mixing matrix, which is parametrized as the product of a lower-triangular matrix and a unitary matrix. Then, we carefully examine the reactor antineutrino flux and conclude that there is no need to modify it from the theoretical aspect. We also calculate the cross sections for neutrino detection in a direct way, together with the complete expressions for the cross sections for the IBD and the elastic antineutrino-electron scattering with a non-unitary mixing matrix. Our analyses separate the production, the oscillation and the detection into three independent parts, as usually treated in the unitary case. The only assumption is that neutrino masses are much smaller than the characteristic energy scales associated with production and detection processes. We provide the cross sections for processes involving NC interactions for each flavor of neutrino (or antineutrino) for the first time. Finally, we evaluate the event rates at TAO and JUNO, as well as exhibiting their capabilities of limiting the non-unitarity parameters with the nominal experimental setups.
	
Since the unitarity violation of the lepton flavor mixing matrix is a natural consequence in certain neutrino mass models, its experimental tests will be the crucial topic for future studies in the era of precision measurements. In general, one can combine the unitarity tests in the long-baseline accelerator neutrino oscillation experiments, such as T2HK~\cite{Hyper-Kamiokande:2018ofw} and DUNE~\cite{DUNE:2015lol}, and the precision electroweak and flavor observables together, in order to perform the global analysis and derive much stronger constraints on non-unitarity parameters. On the other hand, theoretical calculations of physical observables are necessary prerequisites for experimental tests. To match the corresponding sensitivities, we emphasize the importance of carrying out precise calculations within a complete neutrino mass model. In particular, such studies are available now in the type-I seesaw model, based on the derived relations of low-energy observables from original model parameters~\cite{Xing:2023kdj,Xing:2024xwb,Xing:2024gmy}, the systematic investigation of its corresponding low-energy effective theory~\cite{Zhang:2021tsq,Zhang:2021jdf,Wang:2023bdw}, and the one-loop renormalization in the full model~\cite{Huang:2025ubs,Huang:2025qew}. Dedicated studies of unitarity violation in this framework are valuable and will be left for future works.

	\section*{Acknowledgments}
	
The authors are grateful to Prof.~Yu-Feng Li and Prof.~Zhi-zhong Xing for helpful discussions. This work was supported in part by the National Natural Science Foundation of China under grant No.~12475113 and No.~12535007, by the CAS Project for Young Scientists in Basic Research (YSBR-099), and by the Scientific and Technological Innovation Program of IHEP under grant No.~E55457U2. 
	
%	\appendix

\end{document}